\documentstyle[prd,preprint,aps,epsf]{revtex}  
\def\a{\alpha}   \def\b{\beta}           \def\d{\delta}
    \def\f{\phi}       
\def\g{\gamma}   \def\G{\Gamma}        
  \def\m{\mu}              
\def\o{\omega}      \def\p{\psi}       
\def\s{\sigma}       \def\t{\tau}
\def\vf{\varphi}             

\def\CD{{\cal D}}   
   \def\CL{{\cal L}}

\def\slash#1{\,/\kern-7pt#1}
\def\rd{\partial}

\def\darr#1{\raise1.5ex\hbox{$\leftrightarrow$}\mkern-16.5mu #1}

\def\rds{/\kern-6pt\rd}
\newcommand{\be}{\begin{equation}}
\newcommand{\bea}{\begin{eqnarray}}
\newcommand{\ee}{\end{equation}}
\newcommand{\eea}{\end{eqnarray}}
\newcommand{\ba}[1]{\left(\begin{array}{#1}}
\newcommand{\ea}{\end{array}\right)}
\newcommand{\fr}[2]{\frac{#1}{#2}}
\newcommand{\nn}{\nonumber}

\newcommand{\id}{\kern0.2em\rule{0.1mm}{0.71em}
                 \kern0.12em\rule{0.1mm}{0.71em}
                 \kern-0.27em\rule[0.68em]{0.27em}{0.1mm}
                 \kern-0.30em\rule{0.44em}{0.1mm}\rule{0.1em}{-1mm}}

\def\prd#1#2#3{Phys. Rev.  {\bf D{#1}}, {#2} ({#3})}
\def\prs#1#2#3{Phys. Rev. {\bf {#1}}, {#2} ({#3})}

\def\plb#1#2#3{Phys. Lett. {\bf B{#1}}, {#2} ({#3})}

\def\npb#1#2#3{Nucl. Phys. {\bf B{#1}}, {#2} ({#3})}

\def\ap#1#2#3{Ann. Phys. {\bf {#1}}, {#2} ({#3})}
\def\jkps#1#2#3{J. Kor. Phys. Soc. {\bf {#1}}, {#2} ({#3})}
\def\zp#1#2#3{Z. Phys. {\bf {#1}}, {#2} ({#3})}
\def\mpl#1#2#3{Mod.Phys.Lett. {\bf{#1}},{#2},({#3})}

\def\hatf{\hat{\varphi}}

\begin{document}                
\draft
\title{Perturbative Expansion around the Gaussian Effective 
Potential of the Fermion Field Theory}
\author{Geon Hyoung Lee%
        \footnote{email: {\sl ghlee@theory.yonsei.ac.kr}} 
        Tack Hwi Lee%
        \footnote{email: {\sl thlee@theory.yonsei.ac.kr}} 
and
        Jae Hyung Yee%
        \footnote{email: {\sl jhyee@phya.yonsei.ac.kr}}}
\address{Department of Physics and Institute for Mathematical Sciences\\
     Yonsei University\\
     Seoul 120-749, Korea}

\maketitle
\begin{abstract}                
We have extended the perturbative expansion method around the Gaussian effective action
to the fermionic field theory, by taking the 2-dimensional Gross-Neveu model as an
example. We have computed both the zero temperature and the finite temperature effective 
potentials of the Gross-Neveu model up to the first perturbative  correction terms, and 
have found that the critical temperature, at which dynamically broken symmetry is restored,
is significantly improved for the small value of the flavour number.     
\end{abstract}
\newpage
\section{INTRODUCTION}  
One of the most important problems in quantum field theory is to develop 
efficient methods of non-perturbative approximation.
The variational Gaussian approximation method has provided an efficient and
convenient device in obtaining non-perturbative information from various quantum
field theories\cite{gaussian}.
The variational method, however, has a serious shortcoming that it does not 
provide a systematic technique to compute the correction terms to the approximation.
Efforts to establish systematic method to improve the variational approximation
have been made by the authors of Ref. \cite{improved} \cite{cea}.
 
Recently a systematic method of perturbative expansion around the Gaussian
effective action(GEA)\cite{lee} has been developed based on the background field method
\cite{bfm}. This method provides an efficient device to compute the generating functionals
for the one-particle-irreducible Green's functions in perturbation series, whose zeroth order
term is the GEA. For the effective potentials of time-independent systems, the result of 
this method is the same as that of the variational perturbation theory developed by
Cea an Tedesco\cite{cea}. 
It has been shown, for the quantum mechanical anharmonic oscillator, that the perturbative 
correction greatly improves the Gaussian approximation even at the first non-trivial
correction level\cite{lee}.

It is the purpose of this paper to extend the perturbative expansion method around GEA to 
the case of fermionic field theories. In the next section a brief review on background 
field method is given.
We then develop the perturbative expansion method around GEA for the two-dimensional
Gross-Neveu model. In section \ref{finite}, we evaluate the finite temperature
effective potential for the Gross-Neveu model\cite{gn}, and show how the perturbative 
correction improves the critical temperature, at which dynamically broken symmetry is  
restored, from that of the Gaussian approximation. We
conclude with some discussions in the last section.

\section{PERTURBATIVE EXPANSION AROUND THE GAUSSIAN EFFECTIVE ACTION OF THE GROSS-NEVEU MODEL}

To give a brief review on the background field method\cite{bfm}, we start 
from the action
\be
S[\f]=\int d^dx \CL[\f(x),\rd_\m\f(x)],
\ee
in the d-dimensional space-time, where $\f$ can be a bosonic or fermionic field variable.
The generating functional for Green's functions is defined by
\be
<0+|0->^J \equiv  e^{i W}
\equiv \int \CD \vf e^{i S[\vf]+iJ\vf},
\label{gfg}
\ee
where $W[J]$ is the generating functional for the connected Green's functions, $J$ is
the external source, 
and the integral convension, $J \f \equiv \int d^d x J(x) \f(x)$, is used 
in the exponent.
The vacuum expectation value of the field operator in the presence of
external source is defined by
\be
\hatf \equiv <\f(x)>^J = \frac{\d}{\d J(x)} W[J],
\ee
and the effective action is defined by the Legendre transformation,
\be
\G[\hatf] \equiv W[J]-\hatf J. 
\label{wantedeffective}
\ee
The functional derivative of $\G[\hatf]$ with respect to $\hatf$ gives,
\be
\frac{\d}{\d \hatf } \G[\hatf] = -J,
\ee
which is of the same form as the classical equation of motion.

We now introduce a new action $S[\f+B]$ obtained by shifting the 
field $\f$ by a background field $B$.
This new action defines a new effective action $\tilde{\G}[\tilde{\vf},B]$,
where $\tilde{\vf}$ is vacuum expectation value of $\f$ field in the
presence of the background field $B$. One can then show that the effective 
action(\ref{wantedeffective}) can be represented as, 
\be
\G[\hatf]=\tilde{\G}[\tilde{\vf}=0,B]|_{B=\hatf}.
\label{action}
\ee
In other words, the effective action $\G[\hatf]$ can be obtained by summing
 all the one-$\tilde{\vf}$-particle-irreducible diagrams with no external 
$\tilde{\vf}$ lines.
This greatly simplifies the perturbative computation of the effective action.

It has been shown in ref.\cite{lee} that one can rearrange the diagrams
in the effective action (\ref{action}) in such a way that the propagator used in the 
perturbative expansion becomes that of the Gaussian approximation, and the zeroth order
term of the effective action consists of the GEA. 

To extend this method to the case of fermionic 
field theories, we consider the 2-dimensional Gross-Neveu model\cite{gn} described 
by the Langrangian density,
\be
\CL_0 ={\bar{\p}}_{\a}i\rds {\p}_{\a}-g{\bar{\p}}_{\a} {\p}_{\a} 
\s -\frac{1}{2}\s^2,\; \a=1,2,3...N,
\ee 
where ${\p}_{\a}$ is 2-dimensional Dirac field with $\a$ denoting the flavour component.
The generating functional for Green's functions is then given by,
\be
Z[\eta,\bar{\eta}, J]=\int \CD \bar{\p}\CD\p\CD\s e^{i\int d^2x{\CL_0+i\bar{\eta}\p
+i\bar{\p}\eta+i\s J }}.
\ee
We are interested in the effective action as a functional of the vacuum expetation
value of the composite field operator, $<\s>=-g<\bar{\p}\p>$, 
and thus we shift only the composite operator; $\s \rightarrow \s+B$. 
Then the generating functional for the Green's functions in the presence of the 
background field $B$ is given by
\bea
\tilde{Z}[ \eta,\bar{\eta},J, B]
&=& \int \CD \bar{\p} \CD \p \CD\s e^{i\int d^2x{\bar{\p}i\rds \p 
- \frac{1}{2}(\s+B)^2-g\bar{\p} \p (\s+B)+i\bar{\eta}\p+i\bar{\p}\eta+i\s J }}  \nn \\
 &=& e^{-\frac{i}{2}{B}^2} e^{-iB \frac{\d}{i\d J}}
e^{ig\frac{\d}{i\d \eta} \frac{\d}{i \bar{\eta}}
\frac{\d}{i\d J}} \int\CD\bar{\p}\CD\p \CD\s
e^{i\int d^2x{\bar{\p}K^{-1}\p+i\bar{\eta}\p+i\bar{\p}\eta}+i\s J}   \nn  \\
&=& det(K^{-1})det(-iK_B^{\frac{1}{2}})e^{-\frac{i}{2}{B}^2}e^{-iB\frac{\d}{i\d J}}
e^{-g\frac{\d}{\d\eta_a}\frac{\d}{\d\bar{\eta}_a}\frac{\d}{\d J}}
e^{\bar{\eta}K\eta+\frac{1}{2}JK_B J},
\label{zz}
\eea
where 
\bea
K^{-1} &=&-(\rds + igB)  \nn \\
K_B^{-1}&=&-i.
\label{zzx}
\eea
The first three factors of the last line of (\ref{zz}) represent 
the one-loop effective action,
and the remaining factor, upon setting 
$\tilde{\s}=\frac{1}{i} \frac{\d}{\d J}log{\tilde{Z}[ \eta,\bar{\eta},J, B]}=0$,  
gives the higher loop contributions to the effective action.
Note that the fermion propagator in Eq.(\ref{zzx}) already contains the interaction effect
through the background field $B$, which is the reason why the procedure of computing higher
loop contributions is simplified.

In order to rearrange the generating functional so that the functional derivative terms
in Eq.(\ref{zz}) represent the perturbative expansion around GEA, we follow
 the procedure of ref.\cite{lee}.
To do this, we consider the following relation,  
\be
\frac{\d}{\d \eta_a ^x}\frac{\d}{\d \bar{\eta}_a ^x}\frac{\d}{\d J^x}
e^{\bar{\eta}G\eta+\frac{1}{2}J G_B J }
= [G_{aa} ^{xx} +(G\eta)_a ^x (\bar{\eta}G)_a ^x](G_B J)^x 
e^{\bar{\eta}G\eta+\frac{1}{2}JG_B J}.
\label{re}
\ee   
where the repeated index $x$ implies the integration over $x$, and 
$G$ and $G_B$ are arbitrary two-point Green's functions for 
fermionic and bosonic fields, respectively.
Note that this type of functional derivative appears in the expansion of the functional
derivative factor in Eq.(\ref{zz}). Eq.(\ref{re}) contains 
a diagram where an internal line coming
out of a point goes back to the same point(cactus-type diagram) 
which constitues GEA, i.e.,
the first term in the right-hand side of Eq.(\ref{re}).

To extract such cactus-type diagrams out of the perturbative expansion, we 
define the primed functional
derivative as,  
\be
(\frac{\d}{\d \eta_a ^x} \frac{\d}{\d \bar{\eta}_a ^x}
\frac{\d}{\d J^x})'  
\equiv  \frac{\d}{\d \eta_a ^x}
\frac{\d}{\d \bar{\eta}_a ^x}\frac{\d}{\d J^x}+G_{aa} ^{xx}\frac{\d}{\d J^x}.
\label{pro}
\ee
Then the primed derivative acting on the generating functional becomes, 
\be
(\frac{\d}{\d{\eta_a}^x}\frac{\d}{\d{\bar{\eta}_a}^x}
\frac{\d}{\d J^x})'e^{\bar{\eta}G\eta+\frac{1}{2}JG_B J}
=(\bar{\eta}G)_a ^x(G\eta)_a ^x{(G_B J)}^x e^{\bar{\eta}G\eta+\frac{1}{2}JG_B J},  
\ee
which does not contain any cactus-type diagrams. In order to express
the generating functional (\ref{zz}) in terms of the primed 
derivative, we need to find Green's functions $G$ and $G_B$ which satisfy,
\bea
&&e^{-g\frac{\d}{\d\eta_a}\frac{\d}{\d\bar{\eta}_a}\frac{\d}{\d J}}
e^{\bar{\eta}K\eta+\frac{1}{2}JK_B J}  \nn  \\
&&=Ne^{A\frac{\d}{\d J}}e^{-g(\frac{\d}{\d\eta_a}
\frac{\d}{\d\bar{\eta}_a}\frac{\d}{\d J})'}
e^{\bar{\eta}G\eta+\frac{1}{2}JG_B J}. 
\label{new3}
\eea

By using the definition of the primed derivative (\ref{pro}), we easily find $G$,$G_B$,$N$ 
and $A$ that satisfy (\ref{new3}):
\bea
N & = & 1    \nn   \\
A & = & -gG_{aa}  \nn  \\
G & = & K    \nn  \\
G_B & = & K_B,
\label{coef}
\eea
where the new Green's functions $G$ and $G_B$ are the same as those of Eq.(\ref{zzx}).  
The reason why the Green's functions $G$ and $G_B$ are simple in this case is 
that we have introduced the composite operator $\s$, which turns the four fermion interaction
into the three particle interaction.
From Eqs.(\ref{zz}) and(\ref{new3}) one finally obtains the effective action:
\bea
e^{i \G[B]}&=&\tilde{Z}[\eta,\bar{\eta},J, B]\mid_{\tilde{\s}=0}   \nn  \\
&=&det(K^{-1})det(-iK_B^{\frac{1}{2}})e^{-\frac{i}{2}{B}^2}I[B],
\label{det}
\eea
where
\be
I[B]=e^{-(B+gK_{aa})\frac{\d}{\d J}}
e^{-g(\frac{\d}{\d\eta_a}\frac{\d}{\d\bar{\eta}_a}\frac{\d}{\d J})'}
e^{\bar{\eta}K\eta+\frac{1}{2}JK_B J}\mid_{\tilde{\s}=0} .
\label{IB}
\ee
Since the functional $I[B]$ does not contain any cactus-type diagrams, the coeffient
of $I[B]$ in Eq.(\ref{det}) gives rise to the GEA as in the case of 
scalar $\f^4$ theory \cite{lee}. 
Since $I[B]$ can be expanded as a power series in the coupling constant $g$, we have 
the perturbative expansion of the effective action around GEA.

The linear term in the exponent of Eq.(\ref{IB}) generates tadpole diagrams, which does not 
contribute to the effetive action \cite{lee}. Note that Eq.(\ref{IB}) has the same struture 
as the higher order contribution part of (\ref{zz}) except that Eq.(\ref{IB}) involves only
the primed derivative. We can therefore compute the perturbative correction terms to 
GEA using the same procedure  as the conventional background field method,
by using the Feynman rule(in momemtum space),      
\bea
propagators &:& K   = \frac{i}{\g\cdot p+m},\;    m = gB , \nn  \\
            & & K_B = i,  \nn  \\
vertex      &:& -g,   \nn  \\
loop\; integral &:& \int \frac{d^2 p}{2\pi^2}.   
\eea

Thus the perturbative correction, $I[B]$, to GEA consists of  one 
$\tilde{\s}$-particle-irreducible
bubble diagrams with no external lines and without cactus-type diagrams, 
as in the case of ${\f}^4$
theory \cite{lee}. 

Up to the first non-trivial contribution from $I[B]$,the effective action 
for the Gross-Neveu
model can easily be shown to be,  
\be
\Gamma=-iTr\ln (K^{-1})-\frac{1}{2}{B}^2+ig^2(K_B)_{xy}(K_{ca})_{xy}(K_{ac})_{xy},
\label{gag}
\ee
where the first two terms are the GEA and the last term is the first order perturbative 
correction to GEA.

The effective potential is defined by 
\be
V_{eff}[B] \equiv -\fr{\G[B]}{\int d^2x},
\ee
where $B$ is the space-time independent background field.Thus the effective potential of
the Gross-Neveu model, up to the first order perturbative correction term, is given by  
\be
V_{eff}(m) = V_G(m)+V_P(m),   
\label{sum}
\ee
where $V_G$ is the Gaussian effective potential,
\be
V_G  =\frac{1}{2}B^2-N\int \frac{dp}{2\pi} \sqrt{p^2+m^2}  
=\frac{m^2}{2 g^2}+\frac{Nm^2}{4\pi^2} (\ln{\frac{m^2}{\Lambda^2}}-1),  
\label{grn}
\ee 
and $V_P$ is the perturbative correction,
\be
V_P  = \frac{N}{2}g^2 m^2 [\int \frac{dp}{2\pi}\frac{1}{\sqrt{p^2+m^2}}]^2
 =\frac{Ng^2}{8\pi^2} m^2 (\ln{\frac{m^2}{\Lambda^2}})^2,
\ee
with $\Lambda$ being the ultraviolet momentum cutoff.

To extract physical information from the effective potentials $V_G$ or $V_{eff}$,
 we need to renormalize them. We can renormalize the effective potentials by requiring 
the renormalization conditions,
\bea
&&\frac{d^2V}{dm^2}\mid_{m=m_0}=\frac{1}{g_R^2}, \\
&&g \s = g_{R} \s_{R},   
\label{reno}
\eea
where $m_0$ represents a renormalization point.

For the Gaussian effective potential, $V_G$, the renormalization condition becomes 
\be
\frac{d^2V_G}{dm^2}\mid_{m=m_0}=\frac{1}{g_R^2}=\frac{1}{g^2}
+\frac{N}{2\pi}(\ln{ \frac{m_0^2}{\Lambda^2}}+2).
\label{gg}
\ee
Substituting Eq.(\ref{gg}) into (\ref{grn}), one obtains  
the renormalized Gaussian effective potential, 
\be
V_G=\frac{1}{2}\frac{m^2}{g_R^2}+\frac{Nm^2}{4\pi}(\ln{\frac{m^2}{m_0^2}}-3),
\ee
which is equivalent to that of the large-$N$ approximation \cite{gn}.

For the perturbatively corrected Gaussian effective potential,
 $V_{eff}$ of Eq.(\ref{sum}), the
renormalization condition (\ref{reno}) becomes  
\be
\frac{d^2V}{dm^2}\mid_{m=m_0}=\frac{1}{g_R^2}=\frac{1}{g^2}
+\frac{N}{2\pi}(\ln{ \frac{m_0^2}{\Lambda^2}}+2)
+\frac{Ng^2}{4\pi^2}[{\ln{\frac{m_0^2}{\Lambda^2}}}^2+6\ln{\frac{m_0^2}{\Lambda^2}}+4].
\label{rerm}
\ee
Requiring that the renormalized coupling constant $g_R$ be finite,
we find the condition,
\be
g^2=\pi(-1\pm \sqrt{1-\frac{4}{N}})/\ln{\frac{m_0^2}{\Lambda^2}},
\label{Kval}.
\ee
for which the effective potential can be  made finite. The case of the negative
 sign in Eq.(\ref{Kval})
leads to unphysical theory. Taking the case  of positive sign in Eq.(\ref{Kval}), we finally
obtain the renormalized effective potential, 
\be
V=\frac{1}{2}\frac{m^2}{g_R^2} + N\sqrt{1-\frac{4}{N}}\frac{m^2}
{4\pi}(\ln{\frac{m^2}{m_0^2}}-3),
\label{vv}
\ee
which clearly shows that the higher order corrections to the Gaussian 
or the large $N$ approximation.
This shows that our method is meanigful when $N$ is lager than 4 for the Gross-Neveu model.

\section{EFFECTIVE POTENTIAL AT FINITE TEMPERATURE}
\label{finite}

To illustrate how our method improves the Gaussian approximation, we compute 
the first non-trivial perturbative correction term to the Gaussian effective  potential
at finite temperature and evaluate the critical temperature, by using the imaginary time 
formulation of the finite temperature quantum field theory. The generating functional  
for the thermal Green's functions is given by, 
\be
Z_{\b}[J]= N\int_{periodic} \CD \vf  e^{\int^\b_0 dt \int
dx\CL(\vf,\dot{\vf})+\vf J},
\label{finz}
\ee
where $\t$ is the imaginary time with period $\b$, the inverse temperature, 
$\dot{\vf}=\frac{\rd \vf}{\rd \t}$, and
the functional integration is performed only for the periodic field $\vf$. Since the 
finite temperature generating functional (\ref{finz}) is of the same form as the zero-temperature
generating functional (\ref{gfg}) except that it is defined in Euclidean space with periodic
boundery condition, we can evaluate the perturbative expansion by following 
the same procedure as in the last section, except that the Feynman rule is now modified by,
\bea
propagator &:&\frac{i}{\g\cdot p-m}  , \; p_{\m}= (p_0=i\o_n, p_1), 
\; \o_n=(2n+1)\pi\b^{-1}  \nn  \\
loop\;  integration &:& \frac{i}{\b}\sum_{i=-\infty}^{\infty} \int \frac{dp}{2\pi}.
\eea       
One can then find the finite temperature effective potential, up to the first 
perturbative correction,
\be
V_{eff}^\b(\s)=\frac{1}{2}\s^2-2N\int \frac{dp}{2\pi}[\frac{\o}{2}+\frac{1}{\b} \ln(1+e^{-\b\o})]
+\frac{ N g^4 \s^2 }{2}[\int \frac{dp}{2\pi}\frac{1}{\o}(1-\frac{2}{e^{\b\o}+1})]^2,
\label{tt}
\ee 
where $\o=\sqrt{p^2+g^2\s^2}$.

To renormalize the effective potential (\ref{tt}), it is convenient to seperate the 
zero-temperature and the finite temperature parts of the effective potential: 
\be
V_{eff}^\b = V_0 +V_{\b} ,  
\label{yt}
\ee
where
\be
V_0 = \frac{1}{2}\s^2 -2N\int \frac{dp}{2\pi}\frac{\o}{2}+
\frac{g^2 m^2 N}{2}[\int \frac{dp}{2\pi}\frac{1}{\o}]^2 , 
\label{ww} 
\ee
\be
V_{\b} = -2N\frac{1}{\b}\int \frac{dp}{2\pi} \ln(1+e^{-\b\o})
- g^2m^2 N[\int \frac{dp}{2\pi}\frac{1}{\o}]F(\b,m)  
+ \frac{g^2 m^2 N}{2}[F(\b,m)]^2,
\label{zerop}
\ee
with $m=g \s$ and $F(\b, m)$ is defined by 
\be
F(\b,m) \equiv \int \frac{dp}{2\pi}\frac{1}{\o}\frac{2}{e^{\b\o}+1}.
\ee
We note that the zero temperature part of the effective potential (\ref{ww})
 is the same as (\ref{yt})
of the last section, and we need only to renormalize this part of the effective potential.
Thus we have the renormalized effective potential at finite temperature,  
\be
V_{eff}^\b=\frac{1}{2}\frac{m^2}{g_R^2} + N\sqrt{1-\frac{4}{N}}\frac{m^2}
{4\pi}(\ln{\frac{m^2}{m_0^2}}-3) +V_{\b},
\label{pp}
\ee
where $m=g_R \s_R$. Since $F(\b,m)$ is finite, the renormalization condition (\ref{Kval}) 
reduces the finite 
temperature part of the effective potential, $V_{\b}$, to   
\be
V_{\b} = -2N\frac{1}{\b}\int \frac{dp}{2\pi} \ln(1+e^{-\b\o})
-\frac{N}{2\pi} [1-\sqrt{1-\frac{4}{N}}]m^2 F(\b,m).  
\label{ee}
\ee

The symmetry that is broken at zero temperature is restored as temperature increases
beyond the critical temperature.  
It is well known that the critical temperature should be zero 
in 2-dimensional space-time\cite{ma},
while in the large-N\cite{jac} and the Gaussian approximations of the Gross-Neveu model imply 
the non-vanishing critical temperature. To see how the perturbatively improved 
Gaussian approximation improves this result, we now evaluate the critical temperature
from the effective potential (\ref{ee}).

Dynamical symmetry breaking in the Gross-Neveu model is manifested by the fact that
the minimum of the zero temperature effective potential occurs at non-vanishing value
of the composite field, $\s_R=\s_m$, which breaks the symmetry of the classical potential, i.e.,
the symmetry under $\s \rightarrow -\s$. $\s_m$ is determined by,      
\be
\frac{dV_0}{d\s}\mid_{\s=\s_m}=1+ N\sqrt{1-\frac{4}{N}}\frac{g^2}{2\pi}
(\ln{\frac{\s_m^2}{\s_0^2}-2})=0. 
\label{f1}
\ee
As the temperature increases the value of $\s_m(\b)$, at which $V_{eff}^\b$ is minimized,
 decreases.
At the critical temperature, $T_c=\frac{1}{\b_c},\; \s_m(\b)$ vanishes, which implies
 the symmetry restoration.
Thus the critical temperature is determined by
\be    
\frac{dV}{d\s}\mid_{\s_R=\s_m (\b_c)=0}=0,
\label{f2}
\ee
which, together with Eqs. (\ref{pp}) and (\ref{ee}), gives the value of the critical temperature,
\be
T_c = 0.57 g_R \s_m.
\ee
Due to Eq.(\ref{f1}), this can be written as, 
\be
T_c=0.57 g_R\s_0 exp[1-\frac{\pi}{g_R^2}\frac{1}{N \sqrt{1-\frac{4}{N}}}],
\ee
where $\s_0$ represents the renormalization point. 
The critical temperature in the Gaussian approximation can similarly be obtained :
\bea
T_c^{gaussian} &=& 0.57 g_R \s_m^{gaussian}  \nn  \\
               &=& 0.57 g_R  \s_0 exp[1-\frac{\pi}{g_R^2}\frac{1}{N}].
\eea
Although the critical temperature $T_c$ from the perturbatively corrected Gaussian 
approxmation does not vanish, it is smaller than that of the Gaussian approximation.
This shows that the perturbatively improved Gaussian approximation significantly
improves the critical temperature for small $N$, the flavour number of the fermion field  
even at the first order correction level. 

\section{DISCUSSION} 
We have extended the method of the perturbative expansion around GEA developed 
in ref.\cite{lee}
to the fermionic field theory, taking the Gross-Neveu model as an example. 
This method is based on the observation that the Gaussian effective action consists of 
cactus-type diagrams, which is extracted out of the functional derivative part of the effective
action, i.e., the last two factors of the last line of Eq.(\ref{zz}), by 
introducing the primed functional derivative defined in Eq.(\ref{pro}).
This procedure effectively rearranges the diagrams in such a way that the zeroth order
term of the effective action consists of the GEA. Due to the introduction
of the composite field, $\s=-g \bar{\p} \p$, as an order parameter, 
the expansion works only for the flavour number
$N$ larger than 4. 

In the last section the finite temperature effective potential of the Gross-Neveu model
is obtained up to the first non-trivial perturbative correction terms, and the critical
temperature, at which the dynamically broken symmetry is restored, is evaluated.
The result shows a significant improvement of the critical temperature compared to the
Gaussian and the lagre-$N$ results for small value of the flavour number.    
       
\section*{Acknowledgements}
This work was supported in part by the Korea Science and Engineering Foundation 
under Grant No. 95-0701-04-01-3 and 065-0200-001-2, the Center for Theoretical 
Physics (SNU), 
and the Basic Science Research Institute Program, Ministry of Education, 
under project No. BSRI-97-2425.




\newpage

\begin{thebibliography}{99999999}
\bibitem{gaussian} T. Barnes  and G.  I. Ghandour,  
                \prd{22}{924}{1980};
                P. M. Stevenson, 
                \prd{30}{1712}{1984};\prd{32}{1389}{1985};
                S.-Y.Pi and M.Samiullah,
                \prd{36}{3128}{1987};
                O.Eboli, R.Jackiw and S.-Y.Pi,
                \prd{37}{3557}{1988};
                S.K. Kim, W. Namgung, K.S. Soh and J.H. Yee,
                \prd{41}{3792}{1990};\prd{43}{2046}{1991};\ap{214}{142}{1992};
                S.K. Kim, J. Yang, K.S. Soh and J.H. Yee,
                \prd{40}{2647}{1989};
                S. Hyun, G.H. Lee and J.H. Yee,
                \prd{50}{6542}{1994};
                H.-J. Lee, K. Na and J.H. Yee,
                \prd{51}{3125}{1995};
                H.-J. Lee, S. Lee and J.H. Yee, 
                \jkps{29}{5751}{1996} and
                references therein.
\bibitem{improved}   J.M.Cornwall, R.Jackiw, and E. Tomboulis,
               \prd{10}{2428}{1974};
               L. Polley and D.E.L. Pottinger, eds., 
               {\em Variational calculations in quantum field theory},
               World Scientific, Singapore (1987);
                L. Polley and U. Ritschel, 
                  \plb{221}{44}{1989};
                U. Ritschel,
                   \plb{227}{251}{1989}; \zp{47}{457}{1990}; 
                H. Verschelde and M. Coppens,
                   \plb{287}{133}{1992};
                A.Okopi\'{n}ska, 
               \prd{35}{1835}{1987};
               \prd{38}{2507}{1988};
                Ann. Phys.(N.Y.) {\bf 228}, 19 (1993).
              I. Stancu and P. M. Stevenson, 
                 \prd{42}{2710}{1990};
              I. Stancu, 
                 \prd{43}{1283}{1991};
              K.G.Klimenko, \zp{C60}{677}{1993}, \mpl{A9}{1767}{1994}.
\bibitem{cea} H. Kleinert, Phys. Lett. {\bf A173}, 332(1993);
              P. Cea,
                \plb{236}{191}{1990};
              P.Cea and L.Tedesco,
                 \plb{335}{423}{1994};\prd{55}{4967}{1997};
              S.K.You,K.J.Jeon,C.K.Kim,and K.Nahm,Eur.\zp{19}{179}{1998}.
\bibitem{lee} G. H. Lee and J. H. Yee,
              \prd{56}{6573}{1997}.

\bibitem{bfm}  B.S. DeWitt,
                  \prs{162}{1195}{1967};
                G.t' Hooft,
                in Acta Universitatis Wratislavensis no. 38,
                12th Winter School of Theoretical physics in Karpacz;
                Functional and probabilistic methods
                in quantum field theory vol. I (1975);
                L.F.Abbott,
                   \npb{189}{185}{1981};
                C. Lee,
                  \npb{207}{157}{1982};
                W.Ditrich and M. Reuter,
                  {\em Selected Topics in Gauge Theories},
                   Springer-Verlag (1985).
\bibitem{gn} D.J.Gross, and A. Neveu,
              \prd{10}{3235}{1974}.
\bibitem{ma} Roger F. Darshen, Shang-keng Ma, R. Rajaraman,
               \prd{11}{1499}{1975}.
\bibitem{jac} L. Jacobs, 
               \prd{10}{3956}{1974}
\end{thebibliography}
\end{document}